\documentclass[prl,twocolumn,showpacs,preprintnumbers,amsmath,amssymb]{revtex4}

\usepackage{graphicx}
\usepackage{dcolumn}
\usepackage{bm}

\def\k#1{| #1 \rangle}

\def\bk#1#2{\langle #1 | #2 \rangle}

\def\oiint{\int\@oiint\int}
\def\@oiint{%
   \mathchoice
      {\mkern-18mu\bigcirc\mkern-18mu}%
      {\mkern-12.5mu\circ\mkern-12.5mu}%
      {\mkern-12.5mu\circ\mkern-12.5mu}%
      {\mkern-12.5mu\circ\mkern-12.5mu}}%

      \def\band#1{\scriptscriptstyle(#1)}

\begin{document}

\preprint{CPP-metrology-0}

\title{The Cooper Pair Pump as a Quantized Current Source}

\author{R. Leone}
\affiliation{Institut N\'eel, C.N.R.S.- Universit\'e Joseph Fourier, BP 166,
38042 Grenoble Cedex 9, France}
\author{P. Lafarge}
\affiliation{Laboratoire Mat\'eriaux et Ph\'enom\`enes Quantiques,
Universit\'e Paris Diderot - Paris 7, C.N.R.S. UMR 7162, 75205
Paris Cedex 13, France}
\author{L. P. L\'evy}
\affiliation{Institut N\'eel, C.N.R.S.- Universit\'e Joseph Fourier, BP 166,
38042 Grenoble Cedex 9, France}

\date{\today}

\begin{abstract}
A new charge quantization in a phase-polarized Cooper Pair Pump (CPP) is
proposed, based on the topological properties of its Hamiltonian ground state
over a three-dimensional parameter space $\mathbb{P}$.  The charge is quantized
using a set of path in $\mathbb{P}$ covering the surface of a torus, and is a
multiple of the integer Chern index $c_1$ of this surface. This quantization is
asymptotic but the pumped charge converges rapidly to the quantized value with
the increase in the path frequency. The topological nature of the current makes
this CPP implementation an excellent candidate for a metrological current
standard.
\end{abstract}


 \maketitle

Topological defects occur in a number of different settings.  In superfluids or
magnets, the ground state is described by an order parameter. Topological
defects (vortices, skyrmions...) are singular points of the order parameter
field\cite{Mermin79}. A topological charge is assigned to the mapping from
physical space to order parameter space. In quantum systems with few degrees of
freedom, the energy levels and eigenstates may depend on a set of continuous
classical parameters (or band indices).  When the number of parameters is
sufficient (three for a complex Hamiltonian), isolated degeneracies between two
neighboring levels (or bands) can occur.  The presence of a degeneracy has many
physical consequences.  For quantum wavefunctions, the degeneracies can then be
viewed as a defect in their phase field\cite{Berry84}. Some classical examples
are well known in polyatomic molecules\cite{Herzberg63}, where the nuclear
coordinates form a set of semiclassical parameters (within the Born-Oppenheimer
approximation) for the electronic structure. Energy manifolds can have conical
intersections at isolated values of the nuclear
coordinates\cite{Faure00,Zhilinskii01}. When a topological charge is assigned
to these degeneracies, the change of multiplicity of rotation-vibrations levels
as a function of nuclear coordinates are easily understood in topological
terms. In molecular magnets, isolated degeneracies have also been found for
specific directions and values of the magnetic
field\cite{Wernsdorfer99,Bruno06}.  The integer quantum Hall effect has also
been interpreted in term of Chern indices\cite{TKNN82,Kohmoto84} which are the
total topological charge included within a filled Landau band. In all these
examples, a topological charge is associated to an isolated degeneracy between
neighboring bands, where the defect in the wavefunctions phase field occur. Its
physics can be described by a ``monopole'' placed at the degenerate point in
parameter space.  The flux of the magnetic field produced by the monopole
through any surface enclosing the defect is proportional to the Chern index for
this surface. For a the two dimensional electron gas, the Hall conductivity of
a band is $\sigma_{xy}=\frac{e^2}{h}\times c_1$ where $c_1$ is the Chern index
assigned to the band\cite{TKNN82}. The phase change of a wavefunction on a
closed path is proportional to the integral of the potential vector, which is
also the flux through the surface bounded by the path. This phase change,
better known as Berry's phase, takes a particular value ($\pi$) for some of the
physically relevant paths enclosing degeneracies.  In molecular magnets, this
phase change leads to destructive interferences which quench quantum tunnelling
at the degeneracy points.

In this letter, the relevance of this physics to superconducting
circuits is illustrated around an example, the Cooper Pair Pump
(CPP), which lowest two energy levels form two bands which depend
on continuous parameters (gate charges or voltages and magnetic
flux). The parameter space is in this case three dimensional and
degeneracies (defects) occur at isolated points in parameter
space. We demonstrate how a charge quantization, proportional to
the defect topological charge, can be exploited to produce a
current source with metrological accuracy.

\begin{figure}[b]
\centering
\includegraphics[width=0.65\columnwidth]{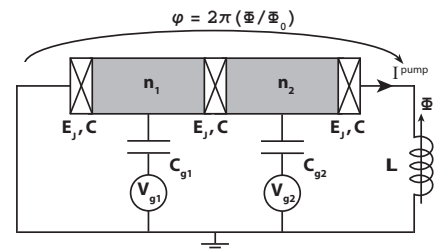}
\caption{The Cooper Pair Pump.}\label{CPP}
\end{figure}
The CPP-circuit represented in Fig.~\ref{CPP} includes a small
inductance in serie with the Cooper pair pump.  The CPP has three
Josephson junctions in series which define two superconducting
islands.  As long as the device is dominated by the Coulomb
charging energy $E_C=\frac{(2e)^2}{4C}$ of the islands, the
topological properties of the circuit do not depend on the precise
value of the Josephson couplings $E_J$ or capacitances $C$, which
can be taken as identical. Let  $n_1$ and $n_2$ be the number of
Cooper pairs in excess on each island and $n_{g1}=\frac{C_{g1}
V_{g1}}{2e}$, $n_{g2}=\frac{C_{g2} V_{g2}}{2e}$, the polarization
charge induced by two gate voltages $V_{g1}$ and $V_{g2}$. The
electrostatic charges on each island are $n_1-n_{g1}$ and
$n_2-n_{g2}$, where $n_{g1}$ and $n_{g2}$ are classical variables
tuning the islands electrostatic energies. The inductance $L$
serves two purposes: the phase $\hat \phi$ across the CPP is
controlled by the magnetic flux $\Phi$ it threads, and it is also
used as the input for the readout circuit measuring the
circulating current through the pump. For a small inductance, the
magnetic energy
$\frac{1}{2L}\left(\hat{\phi}-2\pi\frac{\Phi}{\Phi_0}\right)^2$
defines a sufficiently deep potential minimum: quantum
fluctuations of $\hat \phi$ across the CPP are quenched and the
phase bias is set by the ``classical'' phase
$\varphi=2\pi\frac{\Phi}{\Phi_0}$. The circuit energy can be tuned
through three independent parameters $n_{g1}, n_{g2}$ and
$\varphi$ defining the parameter space $\mathbb{P}$.

\begin{figure}
\centering
\includegraphics[width=\columnwidth]{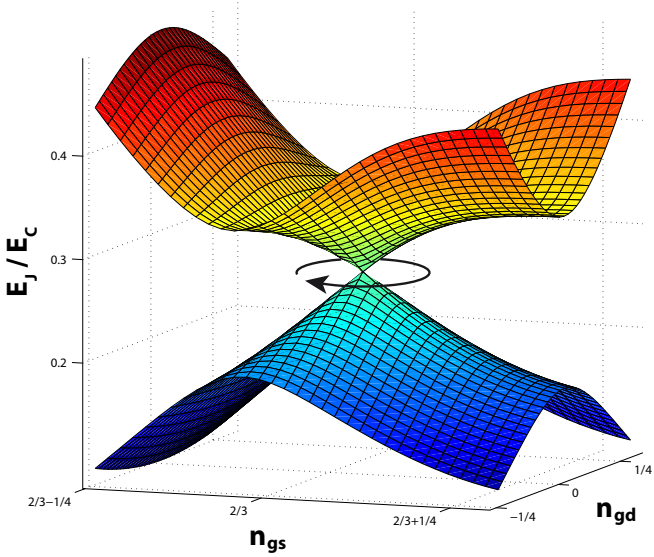}
\caption{(Color) Plot of the energies (in units of the charging energy) of the
two lowest levels as a function of the gate charges $n_{gs}$ and $n_{gd}$ at
$\varphi=\pi$ in the vicinity of the vertex $\mathbf{T_1}$. This ``diabolical
point'' is a conical intersection of the energy sheets between the two lowest
bands. It represents a topological defect in the wavefunction phase fields and
cannot be removed by any perturbations which merely shift its position in
parameter space.} \label{diabolo}
\end{figure}

\begin{figure}[h]
\includegraphics[width=\columnwidth]{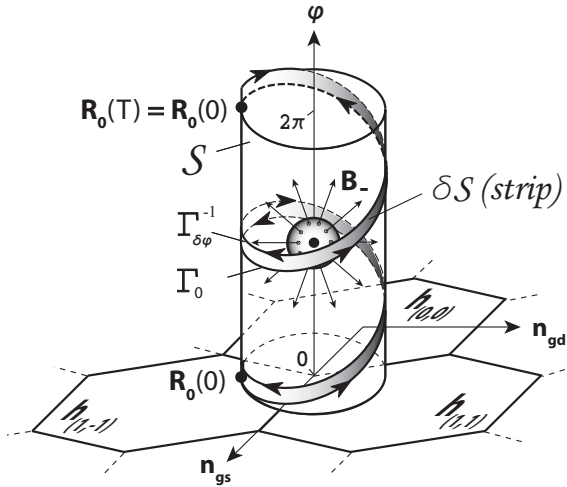}
\caption{Representation of the degeneracy point
$\{T_1,\varphi=\pi\}$ in the parameter space $\mathbb{P}$.  The
cylindrical surface which encloses this point has the topology of
a torus.  The Chern index is defined as the flux through this
surface normalized by $2\pi$. Several paths on this cylinder are
considered in the text.}\label{parameter_space}
\end{figure}

Charging $\mathcal{H}_C$ and Josephson $\mathcal{H}_J$ Hamiltonian
contribute to the CPP energy. It is convenient to write the
charging energy in term of $\hat{n}_s=\hat{n}_1+\hat{n}_2$, the
total charge on the CPP and $\hat{n}_d=\hat{n}_1-\hat{n}_2$, the
charge imbalance between the two islands. Their canonical
conjugates, the phase variables $\hat{\Theta}_s$ and
$\hat{\Theta}_d$ enter in the Josephson Hamiltonian. In term of
$\hat{n}_s$ and $\hat{n}_d$ the charging Hamiltonian is
\begin{equation}
\mathcal{H}_C=E_C\Big[\big(\hat{n}_{s}-n_{gs}\big)^2+\kappa_0\big(\hat{n}_{d}-n_{gd}\big)^2\Big]\,,
\end{equation}
where $E_C=\frac{(2e)^2}{4C}$, $n_{gs}=n_{g1}+n_{g2}$,
$n_{gd}=n_{g1}-n_{g2}$ are the corresponding polarization charges
and $\kappa_0=\frac{1}{3}$ is a capacitance ratio. Each eigenstate
$\k{n_s,n_d}$ is the ground state of $\mathcal{H}_C$ inside an
hexagonal area $h_{(n_s,n_d)}$, centered at the point
$(n_{gs}\!\!=n_s,n_{gd}\!\!=n_d)$, in the $n_{gs}$-$n_{gd}$ plane
(shown as the base plane in Fig.~\ref{parameter_space}). Its
boundaries are electrostatic degeneracy lines between
``neighboring'' charge states, and the vertices are points of
triple degeneracies. The vertices $\{T_1\}=(n_{gs}=\frac{2}{3},
 n_{gd}=0)$ and $\{T_2\}=(n_{gs}=\frac{4}{3}, n_{gd}=0)$ form the unit cell
of this hexagonal lattice. With an appropriate gauge choice, the Josephson
Hamiltonian $\mathcal{H}_J$ depends on the phase bias $\varphi$ as
\begin{equation}
\mathcal{H}_J=-2E_J\cos\hat{\Theta}_s\cos\hat{\Theta}_d-E_{J}\cos(2\hat{\Theta}_d+\varphi)\,.
\end{equation}
The total Hamiltonian is $2\pi$-periodic in $\varphi$.  Lattice
translations in the hexagonal lattice ($n_{gs}$-$n_{gd}$) connect
equivalent but physically different states. Because the parameter
space $\mathbb{P}$ is three dimensional, $\mathcal{H}_J$ lifts the
electrostatic degeneracies except at {\em isolated points} where
the ground level is still degenerate, in $\mathbb{P}$ (a
consequence of the von Neumann-Wigner theorem\cite{vonNeumann29}).
These points occur at the vertices $\{T_1\}$ and $\{T_2\}$ of the
hexagonal lattice for $\varphi=\pi$, as illustrated in
Fig.~\ref{diabolo}. In realistic implementation of the Cooper pair
pump, it is favorable to keep the ratio between Josephson  and
charging energy $\beta=\frac{E_J}{E_C}$ small, an assumption made
throughout this paper.

In the limit of zero Josephson coupling, the electrostatic states
$\k{0,0}, \k{1,1}$ and $\k{1,-1}$ are degenerate at the vertex
$\{T_1\}$. For finite Josephson couplings, the ground state
remains degenerate at this point when $\varphi=\pi$. The two
degenerate ground states at $\mathbf{T_1}=\{T_1,\pi\}$ are
\begin{align}
\k{g_\pm}&=\frac{1}{\sqrt{3}}\bigg[\k{0,0}\mp\frac{\sqrt{3}\mp
1}{2}\k{1,1}\pm \frac{\sqrt{3}\pm 1}{2}\k{1,-1}\bigg]
\end{align}
with energy $E_g(\mathbf{T_1})=\frac{4}{9}\,E_C-\frac{1}{2}\,E_J$ while the
first excited state $\k{e}$ is well separated from the ground state
($E_e=E_g+\frac{3E_J}{2}$). For a small deviation $\delta
\mathbf{R}=\left(\sigma=n_{gs}-\frac{2}{3}, \delta=n_{gd},
\psi=\varphi-\pi\right)$ from degeneracy $\mathbf{T_1}$, a two-level
approximation ($\k{g_\pm}$) is appropriate for small $\beta$. The eigenenergies
$E_\pm(\delta\mathbf{R})=E_g\pm\frac{2}{3}E_C\sqrt{\sigma^2+\frac{\delta^2}{3}+\frac{3}{16}\beta^2\psi^2}
$ display a characteristic conical intersection at $\mathbf{T_1}$. After
rescaling the deviation $\delta \mathbf{R}$ from $\mathbf{T_1}$, as
$b_x=\frac{4}{3}\,E_C\, \sigma, b_y=-\frac{1}{\sqrt{3}}\,E_J\,\psi,
b_z=\frac{4}{4\sqrt{3}}\,\delta$, the two-level Hamiltonian maps onto an
isotropic spin-$\frac{1}{2}$ problem,
\begin{align}
\hat{\mathcal{H}}_{\pm (\mathbf{T_1})}(\mathbf{R})=E_g
\mathbf{1}+\frac{1}{2}\,\boldsymbol{\sigma}\cdot\mathbf{b}(\mathbf{R}).
\end{align}
This makes the conical nature of the intersection explicit
($E_\pm(\delta\mathbf{R})=E_g\pm\frac{|\mathbf{b}|}{2})$, and
spinor eigenstates can be specified using the direction of
$\mathbf{b}$ rather than $\delta\mathbf{R}$.

The topological nature of this degeneracy becomes clear when considering the
adiabatic evolution of quantum states along different closed paths $\Gamma$
``around'' this point. One of them is represented in Fig.~\ref{parameter_space}
as the time evolution of a point $\mathbf{R}(t)$ in $\mathbb{P}$. If
$\k{g_\pm(\mathbf{R}(t))}$ are instantaneous eigenstates of the Hamiltonian,
after a period $T$ around any closed path
$\k{g_\pm(T)}=e^{-i\left(\eta_{\pm}-\gamma_\pm\right)}\k{g_\pm(0)}$ depends on
two phases, $\eta_{\pm}=\frac{1}{\hbar}\int_0^T E_{g_\pm}(\mathbf{R}(t))dt$,
the dynamical phase and $\gamma_\pm=i
\oint_{\Gamma}\mathbf{A_\pm}(\mathbf{R})\cdot d \mathbf{R}$, the geometrical
phase.  It is expressed as an integral of the vector potential
$\mathbf{A}_\pm(\mathbf{R})=\bk{g_\pm(\mathbf{R})}{\boldsymbol{\nabla}g_\pm(\mathbf{R})}$
over the path.  If the closed paths considered are sufficiently close to the
degeneracy $\mathbf{T_1}$ where the two-level approximation is valid,  Berry's
geometrical phase can also be computed using the spin-$\frac{1}{2}$
representation as $\gamma_\pm=i \oint_\Gamma\mathbf{A_\pm}(\mathbf{b})\cdot d
\mathbf{b}$, where the vector potential is azimuthal in this textbook example:
$\mathbf{A}_\pm(\mathbf{b}) =\pm \frac{\cos\theta\mp
1}{2|\mathbf{b}|\sin\theta} \,\mathbf{e}_\phi$ (the Euler angles $\theta,\phi$
specify the direction of $\mathbf{b}$). Its line integral over a closed path
$\Gamma$ is $\gamma_\pm(\Gamma)=\mp \frac{\Omega(\Gamma)}{2}$, where
$\Omega(\Gamma)$ is the solid angle seen from the degeneracy. The vector
potentials $\mathbf{A}_\pm(\mathbf{b})$ can be recognized as the gauge field of
Dirac monopoles of strength $\mp\frac{1}{2}$ placed at $\mathbf{T_1}$.  They
produce a radial magnetic field
$\mathbf{B}_\pm=\mp\frac{\mathbf{b}}{2|\mathbf{b}|^3}$ which flux is $\mp 2\pi$
when integrated over any surface enclosing this single degeneracy. When
normalized to $2\pi$, this flux gives the Chern index $c_1^{\band{\pm}}=\mp 1$
of the surface with respect to the eigenstates $\k{g_\pm}$ and one assignes to
the degeneracy $\mathbf{T_1}$ topological charges $q_{\pm}(\mathbf{T_1})=\mp1$.

This topological picture is not restricted to the immediate neighborhood of
$\mathbf{T_1}$ where a two-level approximation can be made. Let us consider the
cylindrical surface represented in Fig.~\ref{parameter_space}.  Since the top
and bottom planes $\varphi=0$ and $\varphi=2\pi$ are physically equivalent,
this cylinder has the topology of a torus, and its surface is the constant
radius $|\mathbf{R}|=\rho$ cylinder.  The magnetic flux of
$\mathbf{B}_{-}=\boldsymbol{\nabla}\times\mathbf{A}_{-}$ through this closed
surface {\em defines} its Chern index $c_1^{\band{-}}$ of the ground state
$\k{g_-(\mathbf{R})}$\cite{Simon83}. This index is non-zero only if the
topology is non trivial, i.e. if the surface encloses degeneracies. Here the
``cylindrical surface'' encloses only the isolated degeneracy $\mathbf{T_1}$
and Gauss theorem guarantees that its index is identical to the one of the
small sphere around $\mathbf{T_1}$ just considered. For a cylinder around
$\mathbf{T_2}=\{T_2,\pi\}$ the Chern indices have opposite sign and
$q_\pm(\mathbf{T_2})=\pm1$. For a larger surface, the Chern index is the sum of
the topological charges it encloses and when no degeneracies are present, it is
zero.

\begin{figure}
\centering
\includegraphics[width=0.7\columnwidth]{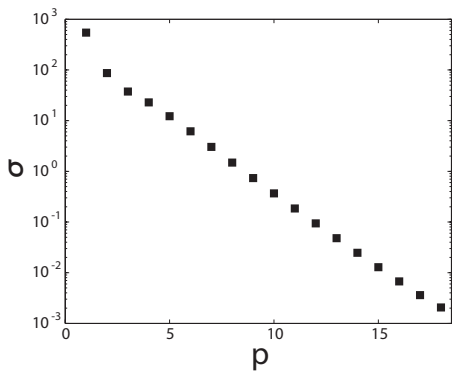}
\caption{Plot of the peak amplitude of $Q^{\rm dyn}(\Gamma_{\varphi_0})$ as a
function of the number of windings $p$ in the interval
$[\varphi_0;\varphi_0+2\pi]$ for a charging energy of $10\:$meV, a Josephson
coupling $E_J=0.5 E_C$ and a period of $10^{-6}\:$sec.}\label{ecarts}
\end{figure}

In a CPP, pumping of charges through the circuit is achieved when
moving the ground state $\k{g_-(\mathbf{R})}$ adiabatically along
a one-dimensional path in $\mathbb{P}$.  In the literature, the
paths considered so far have been closed trajectories around
$\{T_1\}$ or $\{T_2\}$ at constant $\varphi$ for which the charge
transferred is not quantized. {\em When the path is chosen to
cover a surface enclosing one of the singularities, the charge
transferred becomes an integer multiple of the ground state Chern
index}.  This quantization is not only observable but can be
exploited to make a current standard of metrological accuracy, in
a similar way as the quantum Hall effect is a standard of
resistance based on the Chern indices of Landau bands. Let us
consider a helical path $\Gamma_{\varphi_0}$ on the surface of the
cylinder (Fig.~\ref{parameter_space}) starting from the plane
$\varphi=\varphi_0$ and ending in the plane
$\varphi=\varphi_0+2\pi$ while making an integer number $p$ of
windings (in Fig.~\ref{parameter_space}, $\varphi_0=0$ and $p=2$).
The pumped charge $Q^{\rm pump}$ on this path has a dynamical
$Q^{\rm dyn}$ and a geometrical contribution $Q^{\rm geo}$.  The
expressions
\begin{equation}
Q^{\rm dyn}=2e\frac{d\eta_{-}}{d\varphi_0}
(\Gamma_{\varphi_0}),~~Q^{\rm
geo}=-2e\frac{d\gamma_{-}}{d\varphi_0}(\Gamma_{\varphi_0}),
\label{transferred_charge}
\end{equation}
proved elsewere\cite{Leone07}, generalize earlier
results\cite{Aunola03,Mottonen06} to arbitrary three dimensional
closed paths in $\mathbb{P}$.  We first show that the geometrical
contribution is quantized when averaged over the initial phase
$\varphi_0$ of the helix. Taking two helices $\Gamma_{\varphi_0}$
and $\Gamma_{\varphi_0+\delta \varphi}$  shifted in $\varphi$ by
an infinitesimal increment $\delta \varphi$, a closed path
$\Sigma(\varphi_0)$ on the surface of the cylinder (shown in
Fig.~\ref{parameter_space}) can be constructed by adding two small
opposite ``vertical'' segments connecting the paths
$\Gamma_{\varphi_0}$ and $\Gamma^{-1}_{\varphi_0+\delta\varphi}$
which line integrals cancel. Using Eq.~\ref{transferred_charge},
the transferred charge is
\begin{equation}
Q^{\rm
geo}=2e\frac{\gamma_{-}(\Gamma_{\varphi_0})+\gamma_{-}(\Gamma^{-1}_{\varphi_0+\delta\varphi})}{\delta\varphi}
=2e\frac{\gamma_{-}\big(\Sigma(\varphi_0)\big)}{\delta\varphi}\,.
\end{equation}
Using Stokes theorem, the above numerator is also $\oiint_{\delta \mathcal{S}}
{\mathbf B}_{-}(\mathbf{R})\cdot{\hat n}\, dS$, where $\delta \mathcal{S}$ is
the surface of the strip between the two helices (shown in
Fig.~\ref{parameter_space} for $\varphi_0=0$) and $\mathbf{B}_{-}(\mathbf{R}) $
is the ``magnetic'' field produced by the monopole at the degeneracy. When the
initial angle $\varphi_0$ is integrated from $0$ to $2\pi$, adding all the
elementary strips $\delta \mathcal{S}$ generates $p$ times the total surface
$\mathcal{S}$ (Fig.~\ref{parameter_space}). Hence
\begin{equation}
\langle Q^{\rm
geo}\rangle_{\varphi_0}=\frac{2e}{2\pi}\oiint_{\mathcal{S}}{\mathbf
B}_{-}\cdot{\hat n}\, dS=2e\,p\, c_1^{\band{-}}(\mathcal{S})=2e\,p
\end{equation}
is quantized by the Chern index $c_1$ of the cylindrical surface $\mathcal S$
(torus). It is not necessary to average over $\varphi_0$ to reach quantization.
If the number of windings $p$ in $\Gamma$ is large, the path integral of
$\mathbf{A_-}$ over $\Gamma$ reaches the surface integral of $\mathbf B_-$ over
the cylinder and $Q^{\rm geo} \stackrel{p\rightarrow\infty}{=} 2e\, p$. The
increase of accuracy with $p$ being roughly an order of magnitude per
additional winding, this is not only practical but also quite accurate. Using
the same procedures, the dynamical charge $Q^{\rm dyn}$ averages out to zero
because $\eta_-(\Gamma_{\varphi_0})$ is a $2\pi$-periodic function of
$\varphi_0$ since the Hamiltonian is $2\pi$-periodic in $\varphi$. Its
magnitude $\simeq \frac{T}{\hbar}\frac{E_J^3}{E_C^2}$ depends critically on the
ratio $\beta=\frac{E_J}{E_C}$, however the periodic and odd-dependence of
$Q_{\rm dyn}$ with $\varphi_0$ guarantees that its $\varphi_0$-average is zero.
In realistic situations, $\varphi_0$ may not be completely controlled, or the
path may not close perfectly because of flux-noise.  It is therefore
instructive to plot the evolution of the peak amplitude of $Q^{\rm dyn}$ as a
function of the number of winding $p$ in a $\varphi-$period. In
Fig.~\ref{ecarts}, this peak amplitude looses four order of magnitudes when $p$
is increased by $10$, using a ratio $\beta=0.5$ an order of magnitude larger
than its optimal value.  For smaller $\beta$, the decrease is even more
dramatic. Hence, the CPP delivers a very accurate DC current
\begin{align}
I^{\rm pump}=Q^{\rm
pump}\,\nu\stackrel{p\rightarrow\infty}{=}2e\,\nu\,,
\end{align}
where $\nu$ is the winding frequency.

We now consider the possible sources of errors in a practical implementation of
this circuit as a current standard. Low frequency noise in the parameters will
introduce some jitter in the path covered in $\mathbb{P}$: the surface
generated by the path will not be a perfect cylinder. On the other hand, as
long as the jitter amplitude is small compared to the radius, the surface
generated still encloses the topological charge, and no significant error in
quantization should result. The most detrimental source of error comes from
inelastic transitions between the ground state and the first excited state in
the adiabatic evolution, because their topological charges are opposite.  This
means that if the time spent in the first exited state is $\tau$ in a period,
the relative error in the transferred charge will be $-4e\frac{\tau}{T}$. There
are two ways this can happen. For each winding around $T_1$, three saddle
points between the ground and first excited manifolds have to be crossed (along
the hexagonal lines shown in Fig.~\ref{parameter_space}), where the probability
of Landau-Zener tunnelling is largest. For the optimal radius ($\simeq
\frac{1}{3}$) this tunnelling probability is of order $P\simeq
e^{-\left(\frac{3\pi}{2}\right)^2 \frac{E_J}{E_C}\frac{E_J}{h\nu}}$.  To stay
below the part per million level requires that the winding frequency $\nu$ to
stay below $100\:$MHz for a Josephson coupling of $0.05 E_C$, limiting the
pumped current amplitude $2e\nu$ to $32\:$pA. The other source of inelastic
transitions to the first excited state is the back-action of the measuring
device on the CPP, which needs to be rigourously controlled.

The physics discussed in this paper is by no means limited to the CPP: as long
as a circuit has three tunable parameters or more, degeneracies can occur. In
this paper, we emphasized the topological quantization and its strong
robustness to adiabatic parameters fluctuations, a key point for metrological
applications. The realization of this metrological source is under way in our
group: it is a beautiful scientific challenge for fundamental quantum
electronics of real practical value. R. Leone is supported by a fellowship of
the Rh\^{o}ne-Alpes Micro-Nano cluster. The authors are grateful to F. Faure
for enlightening discussions.

\end{document}